\documentclass[conference]{IEEEtran}
\IEEEoverridecommandlockouts
\usepackage{cite}
\usepackage{amsmath,amsfonts}
\usepackage{algorithmic}
\usepackage{graphicx}
\usepackage{textcomp}
\usepackage{xcolor}
\def\BibTeX{{\rm B\kern-.05em{\sc i\kern-.025em b}\kern-.08em
    T\kern-.1667em\lower.7ex\hbox{E}\kern-.125emX}}
\begin{document}

\title{Differing Roles of Leisure and Productivity in GDP - A Machine Learning based comparative analysis of Germany and USA
}

\author{\IEEEauthorblockN{Achintya Ranjan}
\IEEEauthorblockA{\textit{Department of Economics} \\
\textit{Shiv Nadar University}\\
Dadri, India }
\and
\IEEEauthorblockN{Uma Ranjan}
\IEEEauthorblockA{\textit{Faculty of Engineering and Technology} \\
\textit{Sri Ramachandra Institute of Higher Education and Research}\\
Chennai, India 
}
}
\maketitle

\begin{abstract}
In this work, the GDP of a country is modelled as the relative interaction between two agents - working hours, reflecting the social choice of a population, and Total Factor Productivity, reflecting the collective investment in productivity enhancers. It is shown that a Random Forest model can accurately predict the GDP from these two factors. The differences in the choices made by Germany and USA are analysed though Gini importance, SHAP plots and partial dependency. It is shown that the differences in the social structure of the countries are reflected in the relative contribution of working hours and productivity to the GDP.
\end{abstract}


\section{Introduction}

Gross Domestic Product (GDP) is a fundamental measure of economic performance, representing the total monetary value of goods and services produced within a country over a specific period. As a key indicator of economic health, GDP serves as a critical metric for policymakers, economists, and investors, helping guide decisions on resource allocation, fiscal policy, and long-term economic planning. The ability to predict GDP accurately is essential for anticipating economic trends, implementing timely interventions, and mitigating potential downturns.

GDP is typically reported at the end of a fiscal year, but real-time estimation and forecasting have become increasingly important for economic analysis. Short-term GDP forecasts often rely on expenditure-based components such as consumer spending, government expenditures, business investments, and foreign trade. In contrast, longer-term GDP projections emphasize supply-side factors, including capital, labor, and Total Factor Productivity (TFP).

The contribution of capital to GDP growth is measured through investments in physical and intangible assets, such as infrastructure, machinery, and technological advancements. The labor contribution is commonly assessed in terms of total working hours. Meanwhile, TFP measures how efficiently capital and labor are utilized in the production process. TFP encompasses a range of factors, including technological innovation, infrastructure development, and policy frameworks that drive productivity beyond what can be explained by labor and capital alone.

 Recent global trends indicate that countries exhibit different responses to productivity growth. Some nations, like Germany, have leveraged productivity improvements to reduce working hours while sustaining or even increasing GDP. In contrast, the United States has maintained longer working hours alongside productivity gains, leading to higher GDP growth. These patterns reflect underlying social choices regarding work-life balance, economic priorities, and labor policies. However, it remains unclear whether technological advancements and productivity enhancements sufficiently compensate for reduced working hours across all economic contexts. Understanding the relative contributions of labor and TFP to GDP can provide deeper insights into national economic strategies and their impact on growth.
 
This study explores the application of machine learning models for GDP prediction, focusing on the interpretability of feature contributions. Traditional econometric models, such as those based on the Cobb-Douglas production function, often struggle to capture nonlinear relationships and complex interactions between economic variables \cite{mankiw_growth,barro_growth}. Machine learning approaches offer a data-driven alternative that can reveal nuanced patterns in economic behavior. By leveraging explainability metrics such as  feature importance and SHAP (Shapley Additive Explanations), this research aims to provide transparent, interpretable insights into how working hours and TFP influence GDP.

The data for this study is sourced from two different economic databases. Annual GDP and total working hours are obtained from the Organisation for Economic Co-operation and Development (OECD), while TFP data is derived from the Central Reserve Bank of St. Louis.

 The remainder of this paper is structured as follows: Section 2 reviews existing approaches to GDP estimation, Section 3 details the methodology, including data sources and machine learning models, Section 4 presents the results and analysis, and Section 5 concludes with key findings and future research directions.

\section{Related Work}

Traditionally, GDP has been modeled using classical econometric approaches such as the Cobb-Douglas production function and the Solow growth model. The Cobb-Douglas model represents GDP as a function of labor and capital inputs, assuming constant returns to scale and a fixed elasticity of substitution between factors. The Solow model extends this framework by incorporating Total Factor Productivity (TFP) to account for technological progress and efficiency improvements\cite{solow}. While these models offer a strong theoretical foundation, their reliance on predefined functional forms and assumptions about economic relationships limits their adaptability to dynamic economic conditions. In particular, these models assume linearity or log-linearity, making them less effective in capturing nonlinear interactions between economic variables \cite{traditional_gdp_pred,durlauf_growth}.

Beyond production functions, statistical time-series approaches, such as Autoregressive Integrated Moving Average (ARIMA) models and Box-Jenkins methodologies, have been widely used for GDP forecasting \cite{ARIMA_GDP_expenditure,ARIMA_gdp1,box_jenkins_somalia}. These methods rely on historical GDP trends and expenditure components to make predictions. However, they often struggle to incorporate supply-side factors such as productivity and labor dynamics, which are critical to long-term growth analysis. 

In recent years, machine learning (ML) models have gained prominence for economic forecasting due to their ability to capture complex relationships in large datasets \cite{bolivia,india_kaggle_gapminder,bigdata_economics}. Unlike traditional econometric models, ML approaches do not require explicit assumptions about functional forms, making them well-suited for handling nonlinear interactions and heterogeneous economic environments. These models can integrate diverse features, including TFP, working hours, investment rates, and sectoral outputs, leading to more robust and adaptive GDP predictions.

Various ML-based techniques have been employed to improve the accuracy of GDP estimation \cite{GDP_explainable,india_gdp}. Ensemble learning methods, such as Random Forests and Gradient Boosting Machines (GBM), have demonstrated strong predictive performance by aggregating multiple decision trees to reduce overfitting and enhance generalization \cite{rf-gbm}. Time-series models such as MIDAS (Mixed Data Sampling) \cite{midas} have also been explored to incorporate high-frequency indicators into GDP forecasting. More advanced techniques, including deep learning models and evolutionary algorithms like Particle Swarm Optimization (PSO), have further refined GDP predictions by optimizing parameter selection and handling complex dependencies among economic variables \cite{pso_midas,lstm}. Additionally, some studies have leveraged ML models to predict GDP based on social factors, such as demographic trends, consumer sentiment, and internet search behavior \cite{nigeria_social}.

Despite the growing use of ML in economic forecasting, there has been limited work on leveraging machine learning-based feature importance analyses to extract meaningful economic insights. Most studies focus primarily on predictive accuracy rather than interpreting the underlying economic mechanisms driving GDP growth. Additionally, cross-country comparisons of GDP determinants remain scarce, particularly regarding the interplay between labor and productivity in shaping investment behavior.

This study aims to address these gaps by employing a Random Forest model to analyze GDP contributions in Germany and the USA. By using interpretable ML techniques, such as Gini importance and SHAP (Shapley Additive Explanations), this work provides insights into how labor and productivity influence GDP differently across countries. This approach enables a deeper understanding of regional economic dynamics and offers potential explanations for variations in investment patterns and labor market policies.

\section{Proposed Method}

In this work, we propose a machine learning model to understand the relative contributions of working hours and productivity towards GDP. We estimate the GDP using labour hours and Total Factor Productivity using a Random Forest Method. While estimation of GDP typically uses capital, we propose the use of TFP as a surrogate for capital. This is done for two reasons - firstly, capital is not a directly measurable quantity, but in turn is estimated from using the "perpetual inventory method" (PIM), which involves accumulating past investment data (gross fixed capital formation) over time. When estimating capital stock using a perpetual inventory system, potential flaws include: inaccuracy due to potential errors in data entry, lack of consideration for asset depreciation beyond book value, failure to account for asset obsolescence, inability to capture unexpected asset damage or loss, and the need for regular physical inventory checks to verify accuracy; essentially, relying solely on a perpetual system might not fully capture the true state of capital stock due to potential discrepancies between recorded data and physical reality. 
Secondly, advanced economies are primarily driven by exogenous factor such as innovation \cite{tech_productivity},  and capital is only a small part of these exogenous factors. 
The two components of productivity and labour can be viewed as two inputs with GDP being the output. While the amount of labour input in terms of man hours is largely a collective social choice, productivity levels are determined by multiple other factors, including investment in technological advancement.

We look at three explainability factors - the Gini Feature Importance index, the SHAP plots and the partial dependence plots to understand the relative contributions of labor hours and Total Factor Productivity. We present a comparative statistics of the GDP of Germany and USA based on these factors. 

\subsection{Data Sources}

The data for the analysis is drawn from two sources. The annual GDP and Total Working hours are obtained from OECD database. 50 years of data were collected from OECD database  using the OECD data explorer, by searching for the keyword "GDP". From the search results, the data corresponding to Annual GDP and components - expenditure approach, US \$, volume, constant PPPs, reference year 2020, millions was selected. These data are in volume terms and are converted to US dollars using constant Purchasing Power Parities (PPPs). The GDP data was available for various countries for  dfferent periods. Most countries had data from the late 1960's until 2023. 
 
The TFP at constant National Price was taken from the St. Louis Federal Reserve Economic Data (FRED). TFP measures the efficiency with which inputs such as labor or capital are used to produce output in an economy. Hence, the TFP captures the effects of technological progress, innovation and improvements in production processes.  The TFP at constant price level controls for price level differences within a country over time, hence ensuring that changes in TFP reflect real productivity improvements rather than the effects of inflation, deflation or other price fluctuations. 

\subsection{Data Pre-processing}
The TFP data was available only until 2020 for some countries, probably reflecting the impact of COVID-19, when industries were closed, and data collection was not reliable.  Hence, the analysis was restricted to the 50 year period from  1970 to 2020. In addition, all the data were normalized between 0 and 1 using Min Max Scaling. 

\subsection{GDP Model}
Two different ensemble models have been used to predict GDP using TFP and labour hours as inputs. A Random Forest regressor model and a Gradient Bosting Machine with default parameters were used for fitting the GDP. The default parameters allow for arbitrary depth, and improves accuracy. 

\subsection{Explainability framework}

The output of the Random Forest model was analyzed in three different ways of explainability
\begin{itemize}
    \item The Gini feature importance index which indicates the relative contribution of each of the factors to the prediction of the GDP.
    \item SHAP (SHapley Additive exPlanations) plots, which use a game-theoretic approach that measure each factor's contribution to the GDP.
    \item Partial dependency plots which indicates the partial dependence of the target variable (GDP) on the inputs
\end{itemize}

The overall flow of the algorithm is as follows. First, we collect the GDP, TFP and average working hours data over 50 years, and formulate a regression model of GDP. We extract the Gini index to look at the relative weightage of TFP and working hours. We compare this set of indices for two countries, and show that this reflects the well-known social choice of these countries. Partial dependence plots and SHAP indices are additionally used to demonstrate the consistency of explanations. 

\subsection{Economic Implications}
We study the explainability factors from the Random Forest Model to deduce the patterns of investment in USA and Germany, and to quantitatively characterize their collective social choice. 

\section{Results}

\subsection{Model Performance and Data Trends}

The analysis covers a 50-year period from 1970 to 2020 for Germany and the USA. To evaluate the performance of the Random Forest regression model, the dataset for each country was split into training and testing sets, and the predictive accuracy was measured.

\begin{figure}[ht]
\centering
\includegraphics[width=0.4 \textwidth]{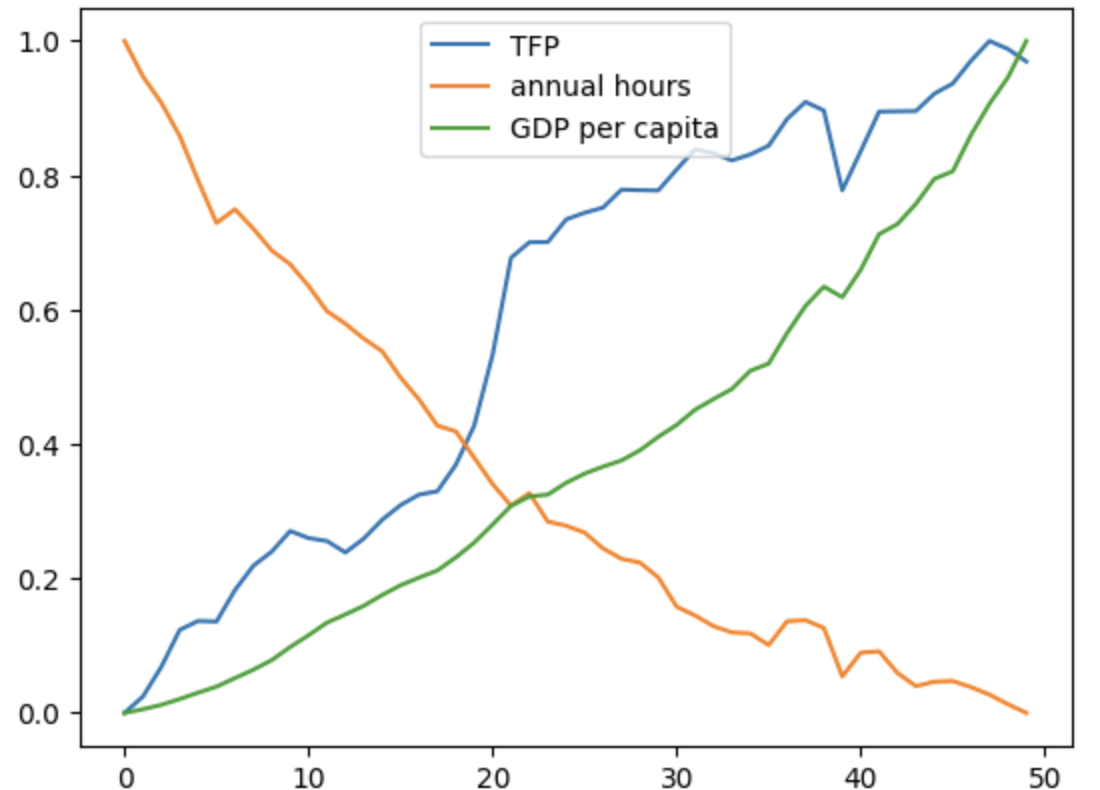}
\includegraphics[width = 0.405\textwidth]{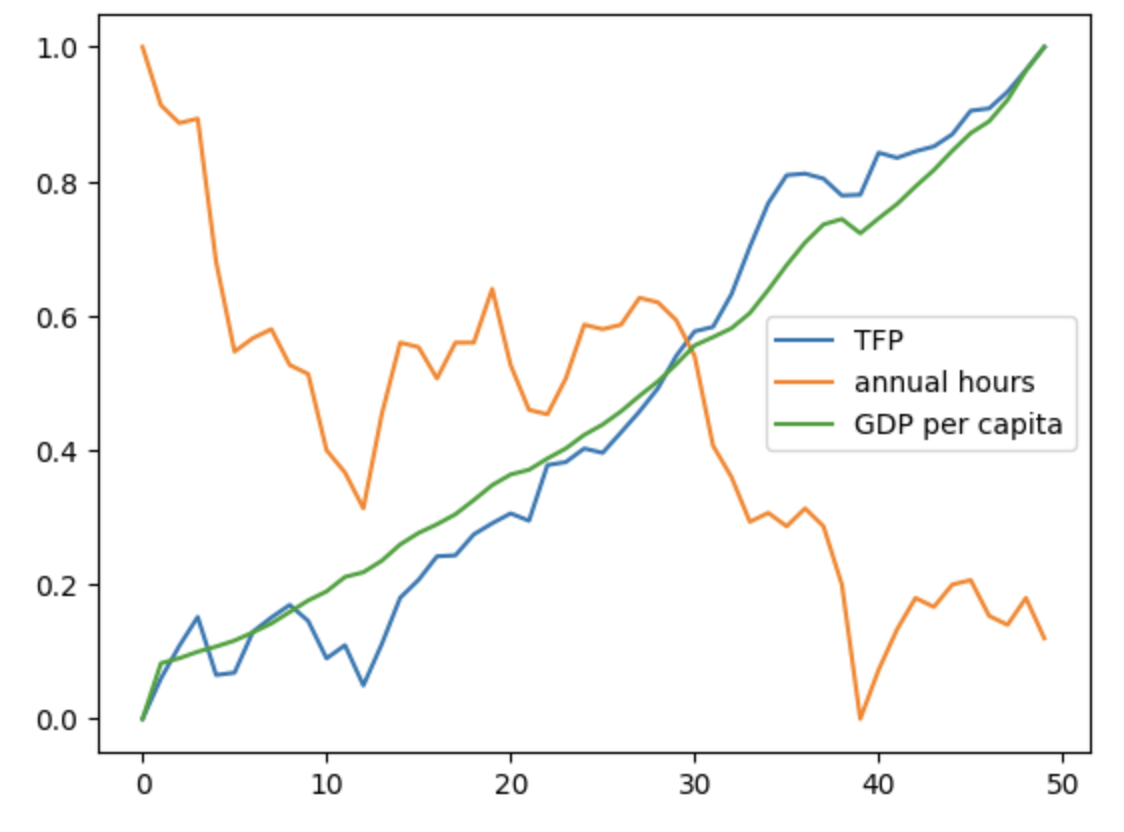}
\caption{Scaled GDP, hours and TFP for (top) Germany and (bottom) USA}
\label{fig:germany_tfp_hours_tfp}
\end{figure}

Figure~\ref{fig:germany_tfp_hours_tfp} presents the normalized GDP, labor hours, and TFP data for both countries over time. As seen in the figure, GDP exhibits a complex dependency on both working hours and TFP, suggesting that a simple linear model may not be sufficient to capture the underlying economic dynamics.

\subsection{Model Fit and Feature Contributions}
\begin{figure}[ht]
\centering
\includegraphics[width=0.4 \textwidth]{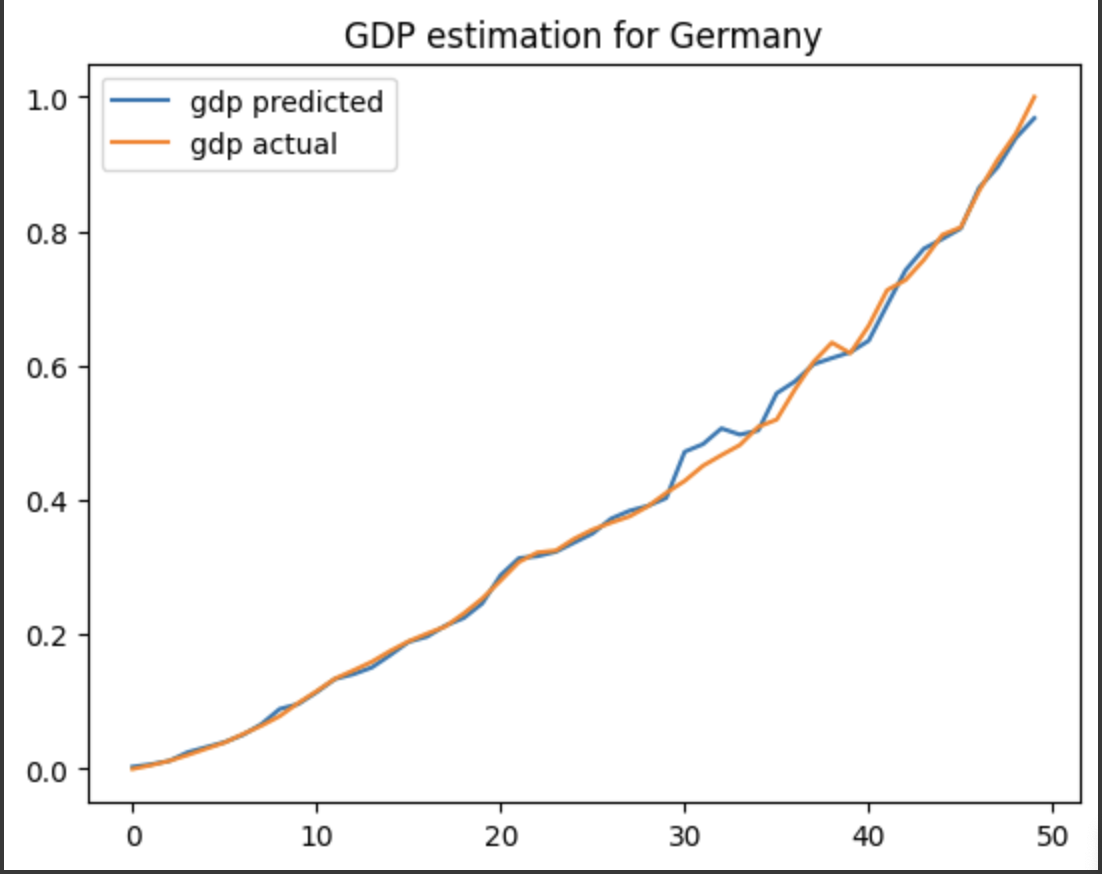}
\includegraphics[width = 0.405\textwidth]{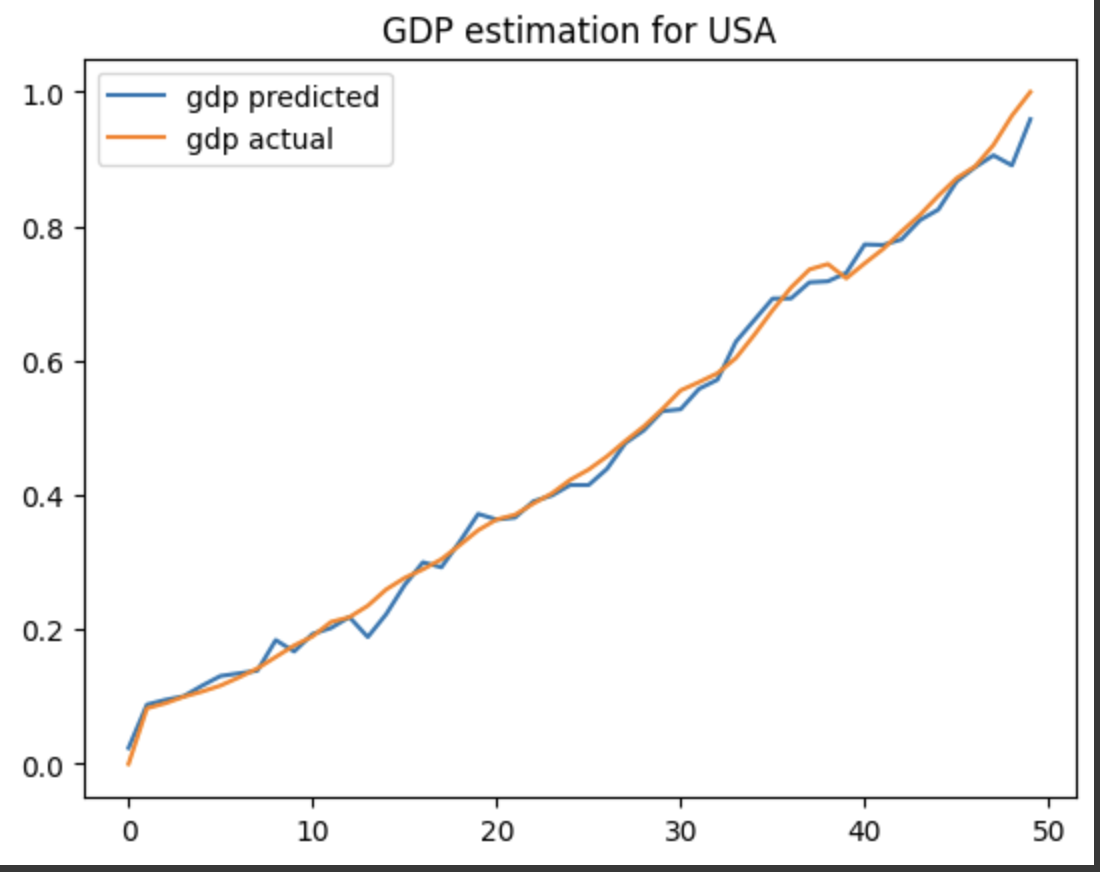}
\caption{Predicted and actual GDP using Random Forest}
\label{fig:gdp_fit}
\end{figure}

The Random Forest model’s predictions closely align with actual GDP values, as shown in Figure~\ref{fig:gdp_fit}. This suggests that the model effectively captures the nonlinear interactions between GDP, working hours, and TFP. There is a period in the early 1990's, where Germany's TFP shows a sharp increase, whereas the GDP entered a period of relative stagnation, followed by a recovery. Economists' analysis of TFP during this period  is divided on the issue, and also cites a possible mis-measurement \cite{burda_reunification,burda_mis}. In Figure~\ref{fig:gdp_fit}, it is seen that the model fails to predict the effect of this sudden disturbance. The data was split into random train-test split. The accuracy of the model is given in Table~\ref{tab:model_accuracy}. The Random Forest model was found to have a mean squared error of 0.0007 and 0.001 for Germany and the USA respectively. The $R^2$ values were found to be 0.989 and 0.981 for Germany and the USA respectively. 

The Gini Feature Importance values, reported in Table~\ref{table:gini_imp}, reveal a stark contrast between the two countries: For Germany, productivity improvements (TFP) contribute approximately 27\% to GDP, indicating a more balanced dependence on both labor hours and productivity whereas for USA, the contribution of TFP is significantly higher, at 95\%, suggesting that the US economy relies predominantly on productivity growth rather than changes in working hours. 

The Gradient Boosting Model reported similar levels of accuracy, with MSE values of 0.0006 and 0.0007 for Germany and the USA respectively, and $R^2$  values of 0.99 and 0.987 for Germany and the USA respectively.

\subsection{SHAP Plot Analysis}

\begin{figure}[ht]
\centering
\includegraphics[width=0.4 \textwidth]{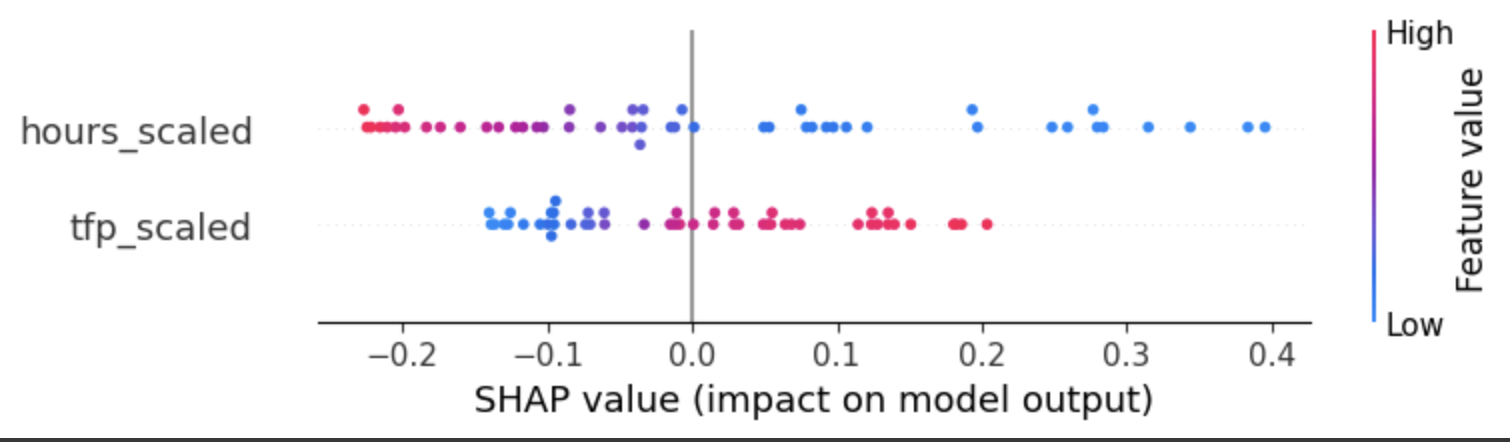}
\includegraphics[width = 0.405\textwidth]{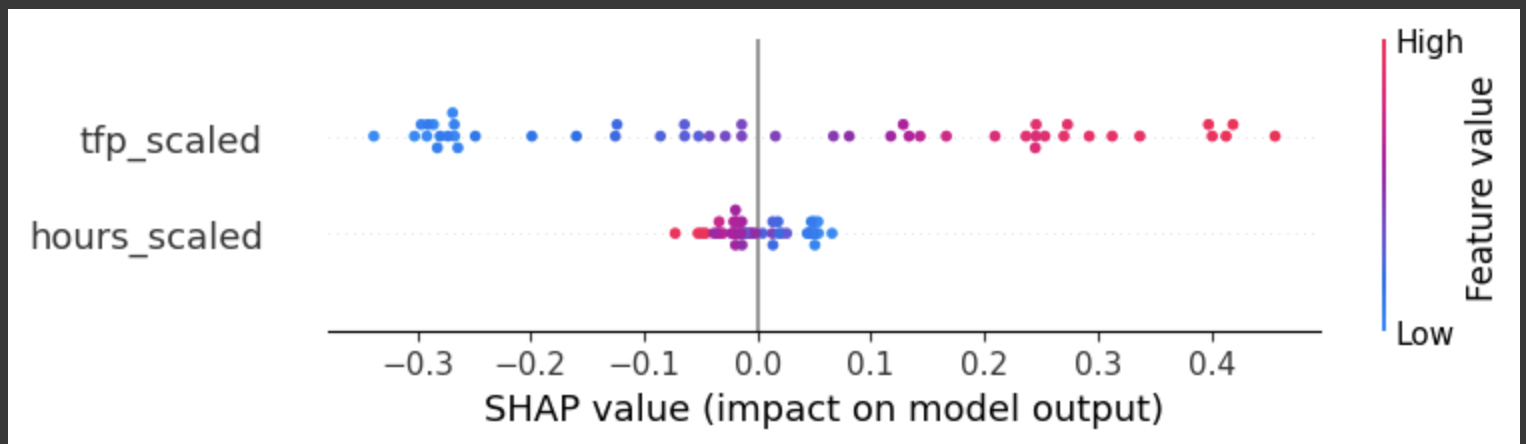}
\caption{SHAP plots for (left) Germany and (right) USA}
\label{fig:shap}
\end{figure}

 Figure~\ref{fig:shap} presents SHAP plots for Germany and the USA, showing how variations in working hours and TFP influence GDP. It is seen that in both countries, low labor hours correlate with high GDP, likely due to the compensatory effect of higher productivity.
Low TFP values are associated with lower GDP, underscoring the importance of technological and productivity improvements.
Germany exhibits a mixed effect of working hours, suggesting that changes in labor hours alone do not consistently impact GDP.
In the USA, the TFP-GDP relationship is nonlinear, with certain levels of TFP producing disproportionately high GDP increases, likely driven by technological advancements and policy interventions.

\subsection{Partial Dependence Analysis}

\begin{figure}[ht]
\centering
\includegraphics[width=0.4 \textwidth]{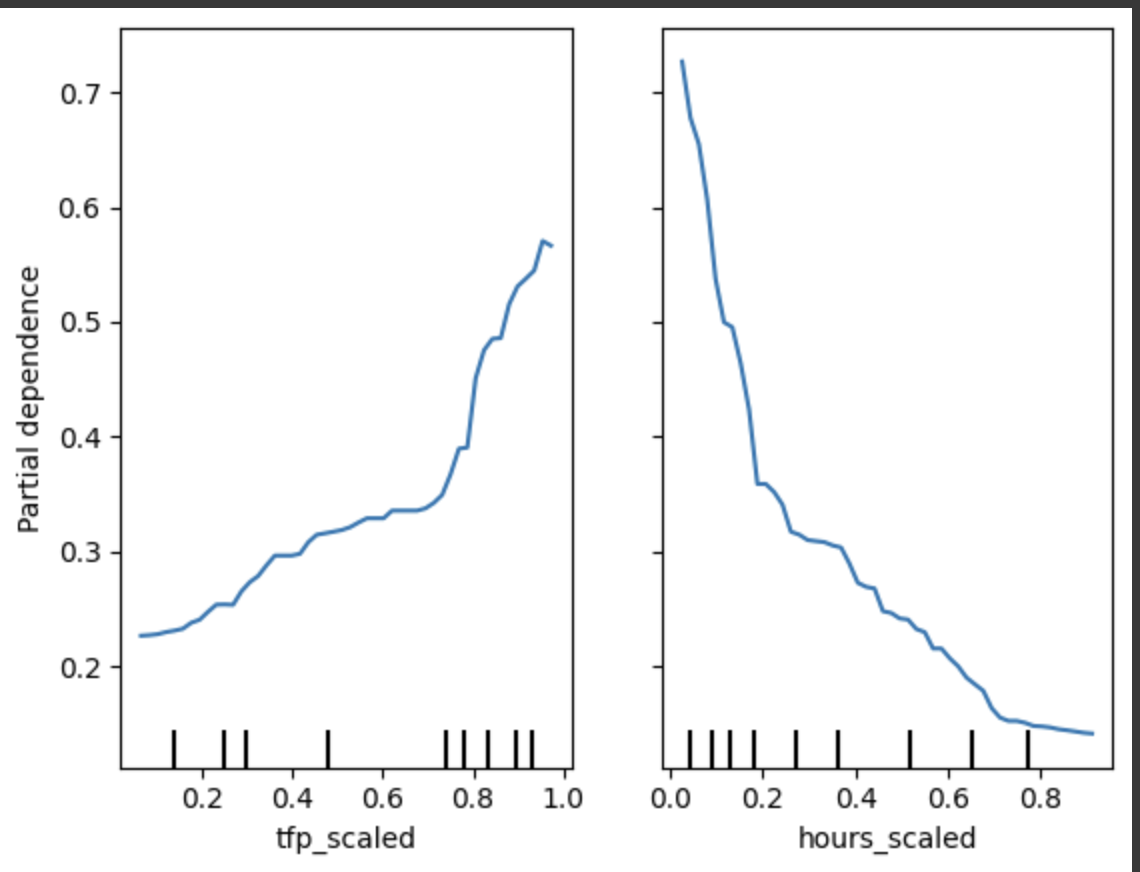}
\includegraphics[width = 0.405\textwidth]{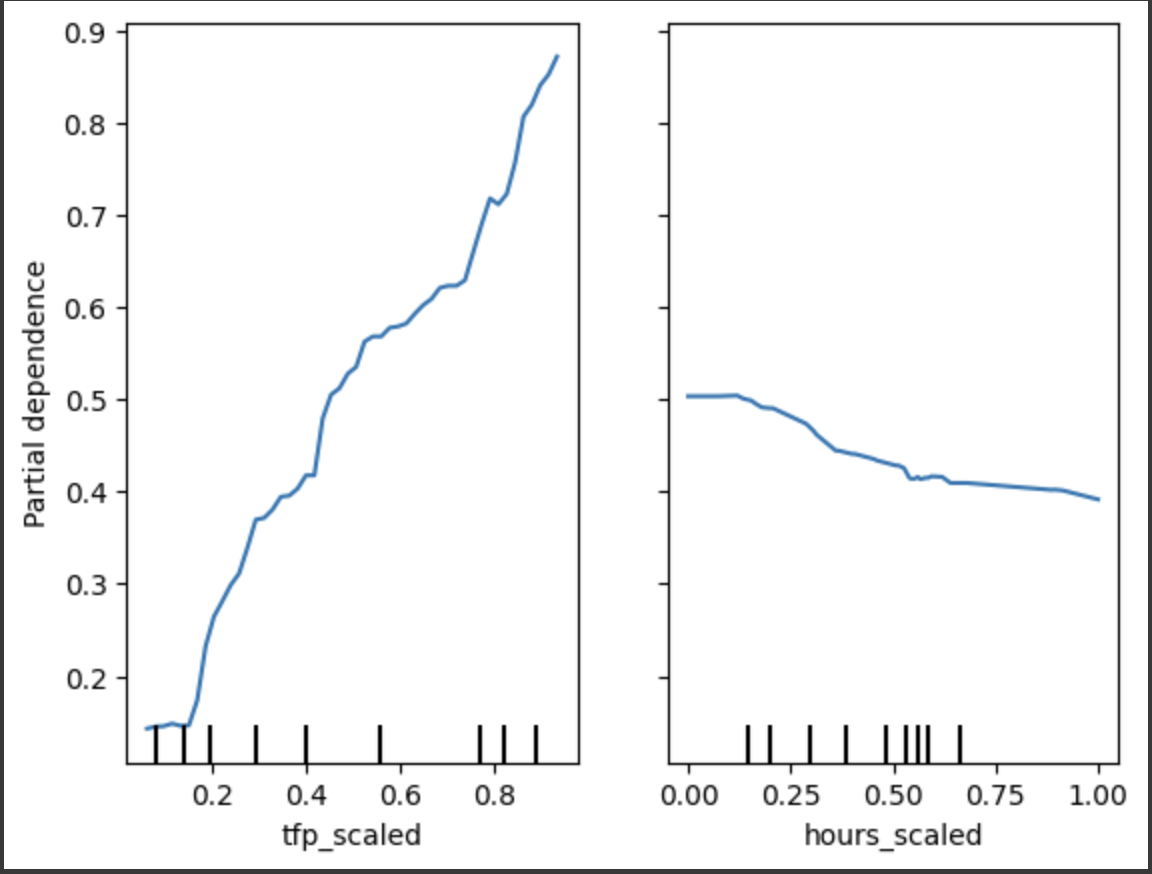}
\caption{Partial Dependence plots for (top) Germany and (bottom) USA}
\label{fig:pdp}
\end{figure}

Figure~\ref{fig:pdp} presents the partial dependence plots, illustrating how GDP responds to changes in TFP and working hours while holding other factors constant. These results reinforce the insights from feature importance analysis:

The USA exhibits a steep dependence on TFP, meaning that even small changes in productivity can lead to substantial GDP shifts.
Germany’s GDP remains more sensitive to working hours, though the effect is still secondary to TFP.
The influence of working hours on GDP in the USA is almost negligible, supporting the idea that productivity-driven growth is the dominant economic strategy.

\subsection{Economic Interpretation and Policy Implications}
These findings highlight fundamental differences in economic structure: Germany follows a more classical economic model, where GDP is still significantly influenced by working hours, despite advances in productivity.
The USA deviates from classical economic patterns, relying almost exclusively on productivity growth to drive GDP expansion. Given that TFP includes factors such as technological advancements, infrastructure, and institutional policies, the results suggest that the USA invests more heavily in technology and productivity-enhancing mechanisms. 

The implications for these are along several directions

\begin{itemize}
\item
{\bf Economic and Policy Explanations for Observed Trends}

The USA’s high dependence on TFP (95\%) suggests that economic growth is driven primarily by technological progress, innovation, and productivity improvements, rather than an increase in labor hours. This is consistent with the country’s long-standing emphasis on automation, R\&D investments, and capital-intensive industries. Moreover, Silicon Valley, financial markets, and a pro-entrepreneurial ecosystem have created a strong innovation-driven economy, where GDP growth is increasingly detached from labor input \cite{tech_productivity}.

Germany, on the other hand, still relies significantly on labor hours in addition to productivity. This is consistent with Germany’s manufacturing-oriented economy, which, despite automation, continues to rely on skilled labor in industries such as automotive and machinery production. Germany’s emphasis on labor protections, work-life balance policies, and collective bargaining agreements may explain why labor hours still play a role in GDP growth. Policy-driven work-hour reductions (e.g., shorter workweeks, vacation laws) may also limit the impact of productivity gains on GDP.

\item
{\bf The Role of Structural Economic Factors}
The Partial Dependence Plots (PDPs) reveal that TFP growth has a steeper effect on GDP in the USA than in Germany, indicating that productivity-driven improvements yield higher economic returns in the USA. This suggests that:

Capital-intensive industries and digital economies in the USA may have stronger productivity spillovers.
Venture capital and startup ecosystems foster rapid technological adoption, further amplifying GDP growth from productivity gains.
Tax policies and corporate structures encourage reinvestment in efficiency-enhancing technologies.
For Germany, the flatter dependence on TFP suggests that productivity improvements alone do not yield as significant GDP gains. This could be attributed to factors such as greater regulation, market rigidity, and workforce-driven policies that may slow the full realization of productivity benefits.

\item
{\bf Labor Hours vs. Productivity: A Non-Uniform Relationship}
A particularly interesting finding is that reducing labor hours does not necessarily lead to lower GDP, particularly in high-TFP economies like the USA. The SHAP analysis reveals that in the USA, lower labor hours often correlate with higher GDP, indicating that productivity gains more than compensate for reductions in working time. This aligns with past research showing that countries with high automation and AI adoption can sustain economic growth despite reducing total labor input.

However, for Germany, the relationship between working hours and GDP is more complex and mixed, suggesting that reductions in labor hours are not always offset by productivity gains. This could mean that in Germany, certain sectors are more sensitive to reductions in labor input, or that structural constraints prevent firms from fully capitalizing on technological advancements.

\item
{\bf Cross-Country Comparisons and Broader Implications}
These findings contribute to the ongoing debate about economic growth models and the sustainability of different approaches. Some key implications include:

Innovation-driven economies (like the USA) can maintain GDP growth with fewer labor inputs by prioritizing R\&D, automation, and efficiency improvements.
Labor-driven economies (like Germany) may need to rethink workforce policies as automation and AI continue to transform industries.
These results also have broader implications for macroeconomic policy, industrial strategy, and workforce planning, as countries navigate the trade-offs between efficiency-driven growth and labor market stability.
\end{itemize}

\begin{table}
\caption{Model Accuracy}
\centering
\begin{tabular}{|c|c|c|}
\hline
Dataset  & MSE & $R^2$ \\
\hline
Germany & 0.0007 & 0.989 \\
USA & 0.001 & 0.981 \\
\hline
\end{tabular}
\hspace*{0.1in}
\vspace*{0.05in}
\begin{tabular}{|c|c|c|}
\hline
Dataset  & MSE & $R^2$ \\
\hline
Germany & 0.0006 & 0.99 \\
USA & 0.0007  & 0.987  \\
\hline
\end{tabular}
Random Forest \hspace{0.7in} Gradient Boosting
\label{tab:model_accuracy}
\end{table}

\begin{table}
\caption{Feature Importance}
\centering
\begin{tabular}{|c|c|}
\hline
Feature & Gini Importance \\
& (Germany) \\
\hline
Hours & 0.727046\\
TFP & 0.272954 \\
\hline
\end{tabular}
\hspace*{0.1in}
\vspace*{0.2in}
\begin{tabular}{|c|c|}
\hline
Feature & Gini Importance \\
& (USA) \\
\hline
Hours & 0.04188 \\
TFP & 0.958113 \\
\hline
\end{tabular}
\label{table:gini_imp}
\end{table}

\section{Conclusion}

This study applies a Random Forest-based machine learning approach to predict GDP using total factor productivity (TFP) and labor hours as key explanatory variables. By leveraging data from Germany and the USA over a 50-year period (1970–2020), the study examines country-specific differences in economic structure and growth dynamics.

The key contributions are:
\begin{itemize}
\item Stark differences between USA and Germany are observed in the role of TFP towards GDP growth
\item Factors like major geopolitical events can cause difficulty in prediction of GDP, or unreliable estimates of data 
\item The relationship between GDP, labor hours, and TFP is highly nonlinear, demonstrating the limitations of traditional econometric models that assume linearity.
\item SHAP and Partial Dependence analyses revealed distinct economic patterns: Germany still follows a classical model where labor hours have a measurable effect on GDP, whereas the USA’s GDP is almost entirely driven by productivity improvements.
\item  Technological investment, infrastructure, and policy interventions are key factors in driving long-term economic growth.
\item The USA’s reliance on productivity growth suggests a shift away from labor-intensive economic expansion, whereas Germany's model suggests a more balanced approach to work hours and technological progress.
\end{itemize}

These insights contribute to the ongoing discourse on economic growth strategies, particularly in the context of technological advancement, automation, and labor market policies. The findings suggest that while reducing working hours does not necessarily harm GDP, sustained economic growth requires strategic investment in productivity-enhancing technologies.

This work also opens up in several directions for future research which can enhance our understanding of GDP dynamics and improve forecasting methodologies:

{\bf Hybrid Machine learning Models}
While Random Forest provided strong predictive accuracy, other machine learning techniques such as gradient boosting (XGBoost, LightGBM), deep learning (LSTMs for time series forecasting), and hybrid models (ML+econometric techniques) could be explored for improved performance and interpretability.

{\bf Causal Inference and Economic Policy Simulations:}

Machine learning models excel at prediction, but understanding causal relationships remains a challenge.
Future research could integrate causal inference techniques (e.g., Granger causality, structural equation modeling) to examine whether productivity growth causes GDP expansion or vice versa.
Policy simulation frameworks could help assess the impact of changes in labor laws, automation adoption, and education investment on GDP growth.

{\bf Sector-Specific Analysis:}
The study currently considers GDP as a whole, but economic growth may be driven by specific sectors (e.g., manufacturing, services, technology, agriculture).
A sector-wise breakdown could reveal whether productivity gains in certain industries drive overall GDP growth more than others.

%GDP prediction typically relies on historical data, but real-time forecasting methods using high-frequency indicators (e.g., Google search trends, financial market movements, satellite imagery) could improve early estimates of economic performance.
%Nowcasting models, which provide real-time GDP estimations using continuously updated data streams, could be explored.

{\bf Ethical and Social Implications:}

The declining role of labor hours in economic growth raises important questions about employment, wage distribution, and income inequality.
Future studies could explore the social consequences of productivity-driven economic models and propose policies that balance economic efficiency with social welfare.

This study highlights the transformative role of machine learning in economic forecasting, demonstrating that data-driven models can capture complex, nonlinear economic interactions that traditional approaches may overlook. By analyzing the interplay between labor, productivity, and GDP, the findings contribute to a more nuanced understanding of economic growth dynamics.

As machine learning models continue to advance, they will play an increasingly important role in shaping economic policy, guiding investment strategies, and informing labor market decisions. However, as the world transitions toward AI-driven economies, policymakers must ensure that productivity-driven growth remains inclusive and sustainable, benefiting society as a whole.


\begin{thebibliography}{1}


\bibitem{traditional_gdp_pred}
Ray C.~Fair: "Evaluating the predictive accuracy of models,
Handbook of Econometrics", Chapter 33,
Elsevier,
Volume 3, 1979-1995
1986

\bibitem{solow} R. Solow, "A Contribution to the Theory of Economic Growth," \textit{Quarterly Journal of Economics}, vol. 70, no. 1, pp. 65-94, 1956.

\bibitem{mankiw_growth} N. G. Mankiw, D. Romer, and D. N. Weil, "A Contribution to the Empirics of Economic Growth," \textit{Quarterly Journal of Economics}, vol. 107, no. 2, pp. 407–437, 1992.

\bibitem{barro_growth} R. J. Barro and X. Sala-i-Martin, \textit{Economic Growth}, Cambridge, MA: MIT Press, 2004.

\bibitem{durlauf_growth} S. N. Durlauf and D. Quah, "The New Empirics of Economic Growth," \textit{Handbook of Macroeconomics}, vol. 1, pp. 235-308, 1999.

\bibitem{ARIMA_gdp1} M. C. Medeiros, G. F. Vasconcelos, A. Veiga, and E. Zilberman, "Forecasting Inflation in a Data-Rich Environment: The Benefits of Machine Learning Methods," \textit{Journal of Business \& Economic Statistics}, vol. 39, no. 1, pp. 98-119, 2021.
\bibitem{ARIMA_GDP_expenditure}
B.~Muma and A.~Karoki:Modeling GDP Using Autoregressive Integrated Moving Average (ARIMA) Model: A Systematic Review. Open Access Library Journal, 9, 1-8, 2022
\bibitem{GDP_explainable}
G.~Maccarrone and G.~Morelli and Sara Spadaccini: "GDP Forecasting: Machine Learning, Linear or Autoregression?", Front. Artif. Intell., volume 4, 2021
\bibitem{india_gdp}
N.~Srinivasan {\it et al}: "Predicting Indian GDP with Machine Learning: A Comparison of Regression Models", 9th International Conference on Advanced Computing and Communication Systems (ICACCS), 1855--1858, 2023
\bibitem{bolivia}
Osmar Bolivar: "GDP nowcasting: A machine learning and remote sensing data-based approach for Bolivia", Latin American Journal of Central Banking, 5 (3),100126, 2024
\bibitem{india_kaggle_gapminder}
A. Thilaka and E. Sundaravalli, "A Machine Learning Approach to GDP Prediction by Analyzing Economic Indicators," 2nd International Conference on Artificial Intelligence and Machine Learning Applications Theme: Healthcare and Internet of Things (AIMLA), Namakkal, India,pp. 1-7, 2024
\bibitem{box_jenkins_somalia}
Mohamed AO.: Modeling and Forecasting Somali Economic Growth Using ARIMA Models. Forecasting.4(4):1038-1050, 2022
\bibitem{bigdata_economics} H. R. Varian, "Big Data: New Tricks for Econometrics," \textit{Journal of Economic Perspectives}, vol. 28, no. 2, pp. 3-28, 2014.
\bibitem{midas}
Vladimir Kuzin, Massimiliano Marcellino, Christian Schumacher:
MIDAS vs. mixed-frequency VAR: Nowcasting GDP in the euro area,
International Journal of Forecasting, 27 (2), 529-542, 2011.
\bibitem{pso_midas} 
Shen F, Yan X, Shang Y.: "A novel hybrid PSO-MIDAS model and its application to the U.S. GDP forecast". PLoS One. 19(12), 2024
\bibitem{lstm}
Hopp, D. Economic Nowcasting with Long Short-Term Memory Artificial Neural Networks (LSTM). Journal of Official Statistics, 38(3), 847-873., 2022 
\bibitem{rf-gbm}
J. Yoon, "Forecasting of Real GDP Growth Using Machine Learning Models: Gradient Boosting and Random Forest Approach". Comput Econ 57, 247–265, 2021.
\bibitem{nigeria_social}
M.D.~Adewale, {\it et al}: "Predicting gross domestic product using the ensemble machine learning method", Systems and Soft Computing, 6, 2024.
\bibitem{tech_productivity} N. Bloom, C. I. Jones, J. Van Reenen, and M. Webb, "Are Ideas Getting Harder to Find?" \textit{American Economic Review}, vol. 110, no. 4, pp. 1104-1144, 2020.
\bibitem{burda_reunification}
M. C. Burda and J.A. Hunt, "From Reunification to Economic Integration: Productivity and the Labor Market in Eastern Germany". Brookings Papers on Economic Activity, 2, 1-92, 2001.
\bibitem{burda_mis}
M.C. Burda, Battista Severgnini,
Total factor productivity convergence in German states since reunification: Evidence and explanations,
Journal of Comparative Economics, 46(1), 192-211, 2018.
%\bibtem{india_gdp}
%N. Srinivasan {\it et al}, "Predicting Indian GDP with Machine Learning: A Comparison of Regression Models," 9th International Conference on Advanced Computing and Communication Systems (ICACCS), Coimbatore, India, pp. 1855-1858, 2023.
\end{thebibliography}
\end{document}